\documentclass[english,reprint]{revtex4-1}
\usepackage[T1]{fontenc}
\usepackage{verbatim}
\usepackage{float}
\usepackage{url}
\usepackage{bm}
\usepackage{amsmath}
\usepackage{amssymb}
\usepackage{graphicx}
\usepackage{setspace}

\usepackage[hidelinks]{hyperref}

\makeatletter
\allowdisplaybreaks

\usepackage{tikz}
\usepackage{amsopn}
\DeclareMathOperator{\tr}{\textrm{tr}}

\def\figwidth{0.95}
\newcommand{\subplotlabel}[2]{
\adjustbox{valign=t}{
\begin{tikzpicture}[font=\sffamily\bfseries]
\draw (0, 0) node[inner sep=0] {
#1
};
\draw (-4.2, 2.4) node {#2};
\end{tikzpicture}
}
}

\usepackage[labelformat=empty,position=top]{subcaption}
\usepackage[export]{adjustbox}
\usepackage{etoolbox}

\def \MyTitle{Estimating Mixture Entropy with Pairwise Distances}
\def \MyAbstract{Mixture distributions arise in many parametric and non-parametric settings---for example, in Gaussian mixture models and in non-parametric 
estimation. It is often necessary to compute the entropy of a mixture, but, in most cases, this
quantity has no closed-form expression,
making some form of approximation necessary. 
We propose a family of estimators based on a pairwise distance function between
mixture components, and show that this estimator class has many attractive properties.
For many distributions of interest,
the proposed estimators are efficient to compute,
differentiable in the mixture parameters, 
and become exact when the mixture components are 
clustered.
We prove this family includes lower and upper bounds on the 
mixture entropy.  The Chernoff $\alpha$-divergence 
gives a lower bound when chosen as the distance function, with the
Bhattacharyya distance providing the tightest lower 
bound for components that are symmetric and members of a location family.
The Kullback--Leibler divergence gives an upper bound when used as the distance
function.
We provide closed-form expressions of these bounds for mixtures of Gaussians,
and discuss their applications to the estimation of mutual information.
We then demonstrate that our bounds are significantly tighter than well-known
existing bounds using numeric simulations.
This estimator class is very useful in optimization problems involving 
maximization/minimization of entropy and mutual information, such as MaxEnt 
and rate distortion problems.}
\def \MyAuthors {
Artemy Kolchinsky $^{1,*}$ and Brendan D. Tracey $^{1,2,\dagger}$
}
\def \MyAffiliations {%
$^{1}$ \quad {Santa Fe Institute, Santa Fe, NM 87501, USA}\\
$^{2}$ \quad {Department of Aeronautics and Astronautics, Massachusetts Institute of Technology, Cambridge, MA 02139, USA}\\
$^{*}$ \quad {artemyk@gmail.com}\\
$^{\dagger}$ \quad {tracey.brendan@gmail.com}
}
\begin{document}
\title{\MyTitle}
\author{ \MyAuthors }
\affiliation{\MyAffiliations }
\begin{abstract}
\MyAbstract
\end{abstract}
\maketitle

\noindent
\textbf{Author's note}: 
\emph{The published version of this manuscript (Kolchinsky and Tracey, \emph{Entropy} 2017, 19(7), 361; doi:10.3390/e19070361) contains a mistake in Section~\ref{sec:Mutual-information-across}. The mistake is corrected in this version of the manuscript.	
}

\section{Introduction}

A mixture distribution is a probability distribution whose density
function is a weighted sum of individual densities. Mixture distributions
are a common choice for modeling probability distributions, in both
parametric settings, for example, learning a mixture of Gaussians statistical
model~\cite{mclachlan2004finite}, and non-parametric settings, such as kernel density estimation. 

It is often necessary to compute the differential entropy~\cite{cover_elements_2012_newversion}
of a random variable with a mixture distribution, which is a measure of the inherent uncertainty
in the outcome of the random variable. Entropy estimation arises in image retrieval
tasks~\cite{goldberger2003efficient}, image alignment and error correction
\cite{viola_empirical_1996}, speech recognition~\cite{hershey_approximating_2007,chen2008accelerated},
analysis of debris spread in rocket launch failures~\cite{capristan2014range},
and many other settings. Entropy also arises in optimization contexts~\cite{schraudolph_optimization_1995,viola_empirical_1996,schraudolph_gradient-based_2004,shwartz_fast_2005},
where it is minimized or maximized under some constraints (e.g., MaxEnt
problems). Finally, entropy also plays a central role in minimization
or maximization of mutual information, such as in problems related
to rate distortion~\cite{kolchinsky_nonlinear_2017}.

Unfortunately, in most cases, the entropy of a mixture distribution
has no known closed-form expression~\cite{contreras2016bounds}. This is true even when the entropy
of each component distribution does have a known closed-form expression.
For instance, the entropy of a Gaussian has a well-known form, while
the entropy of a mixture of Gaussians does not~\cite{carreira2000mode}. As a result, the problem
of finding a tractable and accurate estimate for mixture entropy has
been described as ``a problem of considerable current interest and
practical significance''~\cite{zobay2014variational}. 

One way to approximate mixture entropy is with Monte Carlo (MC) sampling.
MC sampling provides an unbiased estimate of the entropy, and
this estimate can become arbitrarily accurate by increasing the number
of MC samples. Unfortunately, MC sampling is very computationally
intensive, as, for each sample, the (log) probability of the sample
location must be computed under every component in the mixture. MC
sampling typically requires a large number of samples to estimate
entropy, especially in high-dimensions. Sampling is thus typically
impractical, especially for optimization problems where, for every
parameter change, a new entropy estimate is required. Alternatively, it is possible 
to approximate entropy using numerical integration,
but this is also computationally expensive and limited to low-dimensional applications~\cite{beirlant1997nonparametric,joe_estimation_1989}.

Instead of Monte Carlo sampling or numerical integration, one may use an analytic estimator
of mixture entropy. Analytic estimators have estimation bias but
are much more computationally efficient. There are several existing
analytic estimators of entropy, discussed in-depth below. To summarize,
however, commonly-used estimators have significant drawbacks: they
have large bias relative to the true entropy, and/or they are invariant
to the amount of ``overlap'' between mixture components. For example,
many estimators do not depend on the locations of the means in a Gaussian
mixture model.

In this paper, we introduce a novel family of estimators for the mixture
entropy. Each member of this family is defined via a pairwise-distance
function between component densities. The estimators in this family
have several attractive properties. They are computationally efficient,
as long as the pairwise-distance function and the entropy of each
component distribution are easy to compute. The estimation bias of
any member of this family is bounded by a constant. The estimator
is continuous and smooth and is therefore useful for optimization
problems. In addition, we show that when the Chernoff $\alpha$-divergence
(i.e., a scaled R{\'e}nyi divergence) is used as a pairwise-distance
function, the corresponding estimator is a lower-bound on the mixture entropy.
Furthermore, among all the Chernoff $\alpha$-divergences, 
the Bhattacharyya distance ($\alpha=0.5$) provides the tightest lower
bound when the mixture components
are symmetric and belong to a location family (such as a mixture of
Gaussians with equal covariances). We also show that when the Kullback--Leibler
{[}KL{]} divergence is used as a pairwise-distance function, the corresponding
estimator is an upper-bound on the mixture entropy. Finally, our family of estimators
can compute the exact mixture entropy when the component distributions
are grouped into well-separated clusters, a property not shared by
other analytic estimators of entropy. In particular, the bounds mentioned
above converge to the same value for well-separated clusters.

The paper is laid out as follows. We first review mixture distributions
and entropy estimation in Section \ref{sec:Background-and-definitions}. We
then present the class of pairwise distance estimators in Section \ref{sec:Pairwise-Estimators},
prove bounds on the error of any estimator in this class, and show
distance functions that bound the entropy as discussed above. In Section \ref{sec:Entropy-of-Gaussian},
we consider the special case of mixtures of Gaussians, and give explicit
expressions for lower and upper bounds on the mixture entropy. When
all the Gaussian components have the same covariance matrix, we show
that these bounds have particularly simple expressions. In Section \ref{sec:Mutual-information-across},
we consider the closely related problem of estimating the mutual information
between two random variables, and  show that our estimators 
can be directly used to estimate and bound the mutual information.
For the Gaussian case, these can be used to bound the mutual
information across a type of Additive White Noise Gaussian channel.
Finally, in Section \ref{sec:Numerical-results}, we run numerical experiments
and compare the performance of our lower and upper bounds relative
to existing estimators. We consider both mixtures of Gaussians and
mixtures of uniform distributions.

\section{Background and Definitions\label{sec:Background-and-definitions}}

We consider the differential entropy of a continuous random variable $X$, defined as 
\[
H(X):=-\int p_{X}(x)\ln p_{X}(x)\,dx\,,
\]
\vspace{-6pt}
where $p_{X}$ is a mixture distribution,
\[
p_{X}(x)=\sum_{i=1}^{N}c_{i}p_{i}(x)\,,
\]
and where $c_{i}$ indicates the weight of component $i$ ($c_{i}\ge0$,
$\sum_{i}c_{i}=1$)  and $p_{i}$ the probability density of component
$i$.

We can treat the set of component weights as the probabilities of 
outcomes $1 \dots N$ of a discrete random variable $C$, where $\text{Pr}(C=i)=c_{i}$.
Consider the mixed joint distribution of the discrete random variable
$C$ and the continuous random variable $X$,
\[
p_{X,C}(x,i)=p_{i}(x)c_{i}\,,
\]
and note the following identities for conditional and joint entropy~\cite{nair2006entropy},
\[
H\left(X,C\right)=H\left(X\vert C\right)+H\left(C\right)=H\left(C\vert X\right)+H\left(X\right),
\]
where we use $H$ for discrete and differential entropy interchangeably.
Here, the conditional entropies are defined as
\begin{align*}
H(X\vert C) &=\sum_{i}c_{i}H(p_{i}) \,, \\
 H(C\vert X) & =\int p_{X,C}(x,i)\log\frac{p_{X,C}(x,i)}{p_{X}(x)}\,dx\,.
\end{align*}

Using elementary results from information theory~\cite{cover_elements_2012_newversion},
$H(X)$ can be bounded from below by
\begin{equation}
H(X)\ge H(X\vert C)\:,\label{eq:conditional}
\end{equation}
since conditioning can only decrease entropy. Similarly, $H(X)$ can
be bounded from above by 
\begin{equation}
H(X)\le H(X,C)=H(X\vert C)+H(C)\,,\label{eq:joint}
\end{equation}
following from $H(X)=H(X,C)-H(C\vert X)$ and the non-negativity of
the conditional discrete entropy $H(C\vert X)$. This upper bound
on the mixture entropy was previously proposed by Huber et al.~\cite{huber2008entropy}. 

It is easy to see that the bound in Equation (\ref{eq:conditional}) is tight
when all the components have the same distribution, since then $H(p_{X})=H(p_{i})$
for all $i$. The bound in Equation (\ref{eq:joint}) becomes tight when $H(C\vert X)=0$,
i.e., when any sample from $p_{X}$ uniquely determines the component
identity $C$. This occurs when the different mixture components have
non-overlapping supports, $p_{i}(x)>0\implies p_{j}(x)=0$ for all
$x$ and $i\ne j$. More generally, the bound of  Equation (\ref{eq:joint}) becomes
increasingly tight as the mixture distributions move farther apart
from one another.

In the case where the entropy of each component density, $H(p_{i})$ for 
$i=1\dots N$, has a simple closed form
 expression, the bounds in Equations (\ref{eq:conditional})
and (\ref{eq:joint}) can be easily computed. However, neither bound
depends on the ``overlap'' between components. For instance, in a
Gaussian mixture model, these bounds are invariant to changes in the
component means. The bounds are thus unsuitable for many problems;
for instance, in optimization, one typically tunes parameters to
adjust component means, but the above entropy bounds remain the 
same regardless of mean location.

There are two other estimators of the mixture entropy that should
be mentioned. The first estimator is based on kernel density estimation
\cite{joe_estimation_1989,hall1993estimation}. It estimates the entropy
using the mixture probability of the component means, $\mu_{i}$,
\begin{equation}
\hat{H}_{\text{KDE}}(X):=-\sum_{i}c_{i}\ln\sum_{j}c_{j}p_{j}(\mu_{i})\,.\label{eq:h-kde}
\end{equation}

The second estimator is a lower bound that is derived using Jensen's
inequality~\cite{cover_elements_2012_newversion}, $-\text{E}\left[\ln f\left(X\right)\right]\ge-\ln\left[\text{E}(f(X))\right]$,
giving 
\begin{multline}
H(X) :=-\int\sum_{i}c_{i}p_{i}(x)\ln\sum_{j}c_{j}p_{j}(x)\,dx\, \\
\ge-\sum_{i}c_{i}\ln\sum_{j}c_{j}\left(\int p_{i}(x)p_{j}(x)\,dx\right)=:\hat{H}_{\text{ELK}}\,.\label{eq:lowerbound-ELK}
\end{multline}

In the literature, the term $\int p_{i}(x)p_{j}(x)\,dx$ has been
referred to as the ``Cross Information Potential''~\cite{principe2000information,xu2008reproducing}
and the ``Expected Likelihood Kernel''~\cite{jebara2003bhattacharyya,jebara2004probability}
 (ELK, we use this second acronym to label this estimator). When
the component distributions are Gaussian, $p_{i} =\mathcal{N}(\mu_{i},\Sigma_{i})$,
the ELK has a simple closed-form expression,
\begin{equation}
\hat{H}_{\text{ELK}}(X) = -\sum_{i}c_{i}\ln\sum_{j}c_{j} q_{j,i}(\mu_i) \,,\label{eq:huberlower}
\end{equation}
where each $q_{j,i}$ is a Gaussian defined as $q_{j,i} := \mathcal{N}(\mu_{j},\Sigma_{i}+\Sigma_{j})$.
This lower bound was previously proposed for Gaussian mixtures in
\cite{huber2008entropy} and in a more general context in~\cite{contreras2016bounds}.

Both $\hat{H}_{\text{KDE}}$, Equation (\ref{eq:h-kde}),
and $\hat{H}_{\text{ELK}}$, Equation (\ref{eq:huberlower}), are computationally efficient, continuous
and differentiable, and depend on component overlap, making them suitable
for optimization. However, as will be shown via numerical experiments (Section 
\ref{sec:Numerical-results}), they exhibit significant underestimation
bias. At the same time, we will show that for Gaussian mixtures with equal
covariance, $\hat{H}_{\text{KDE}}$ is only an additive constant away from
an estimator in our proposed class.

\section{Estimators Based on Pairwise-Distances\label{sec:Pairwise-Estimators}}
\unskip
\subsection{Overview}

Let $D(p_{i}\Vert p_{j})$ be some (generalized) distance function
between probability densities $p_{i}$ and $p_{j}$. Formally, we
assume that $D$ is a \emph{premetric}, meaning that it is non-negative
and $D(p_{i}\Vert p_{j})=0$ if $p_{i}=p_{j}$. We do not assume that
$D$ is symmetric, nor that it obeys the triangle inequality, nor
that it is strictly greater than 0 when $p_{i}\ne p_{j}$.

For any allowable distance function $D$, we propose the following
entropy estimator:
\begin{equation}
\hat{H}_{D}(X):=H(X\vert C)-\sum_{i}c_{i}\ln\sum_{j}c_{j}e^{-D(p_{i}\Vert p_{j})}\,.\label{eq:pairwise}
\end{equation}

This estimator can be efficiently computed if the entropy of each
component and $D(p_{i}\Vert p_{j})$ for all $i,j$ have simple closed-form
expressions. There are many distribution-distance function pairs that
satisfy these conditions (e.g., Kullback--Leibler divergence, Renyi
divergences, Bregman divergences, f-divergences, etc., for Gaussian,
uniform, exponential, etc.)~\cite{banerjee2005clustering,cichockisimilarity,cichocki2010families,gil2013renyi,crooks2017techreport}.

It is straightforward to show that for any $D$, $\hat{H}_{D}$ falls
between the bounds of  Equations (\ref{eq:conditional}) and (\ref{eq:joint}),
\begin{equation}
H(X\vert C)\le\hat{H}_{D}(X)\le H(X,C)\,.\label{eq:cond-joint-bounds}
\end{equation}

To do so, consider the ``smallest'' and ``largest'' allowable
distance functions, 
\begin{align}
D_{\text{min}}(p_{i}\Vert p_{j}) & =0 \nonumber
\\
D_{\text{max}}(p_{i}\Vert p_{j}) &=\begin{cases}
0, & \text{if }p_{i}=p_{j},\\
\infty, & \text{otherwise.}
\end{cases}\label{eq:dmin-dmax}
\end{align}

For any $D$ and $p_{i},p_{j}$, $D_{\text{min}}(p_{i}\Vert p_{j})\le D(p_{i}\Vert p_{j})\le D_{\text{max}}(p_{i}\Vert p_{j})$,
thus
\[
\hat{H}_{D_{\text{min}}}(X)\le\hat{H}_{D}(X)\le\hat{H}_{D_{\text{max}}}(X).
\]

Plugging $D_{\min}$ into Equation (\ref{eq:pairwise}) (noting that $\sum_{j}c_{j}=1$)
gives $\hat{H}_{D_{\text{min}}}(X)=H(X\vert C)$, while plugging $D_{\text{max}}$
into Equation (\ref{eq:pairwise}) gives 
\begin{multline}
\hat{H}_{D_{\text{max}}}=H\left(X\vert C\right)-\sum_{i}c_{i}\ln\Big(c_{i}+\sum_{j\ne i}c_{j}e^{-D_{\text{max}}(p_{i}\Vert p_{j})}\Big) \\
\le H\left(X\vert C\right)-\sum_{i}c_{i}\ln c_{i}=H(X,C).
\end{multline}

These two inequalities yield Equation (\ref{eq:cond-joint-bounds}). The true
entropy, as shown in Section \ref{sec:Background-and-definitions}, also obeys
{$H\left(X\vert C\right)\le H\left(X\right)\le H\left(X,C\right)$.}
The magnitude of the bias of $\hat{H}_{D}$ is thus bounded by 
\begin{align*}
\left|\hat{H}_{D}(X)-H(X)\right| & \le H\left(X,C\right)-H\left(X\vert C\right)=H(C)\,.
\end{align*}

In the next two subsections, we improve upon the bounds suggested
in Equations (\ref{eq:conditional}) and (\ref{eq:joint}), by examining bounds
induced by particular distance functions.

\subsection{Lower Bound\label{subsec:Lower-Bound}}

The ``Chernoff $\alpha$-divergence''~\cite{nielsen_chernoff_2011,crooks2017techreport}
for some real-valued $\alpha$ is defined as
\begin{equation}
C_{\alpha}(p\Vert q):=-\ln\int p^{\alpha}(x)q^{1-\alpha}(x)\,dx\,.\label{eq:chernoff-divergence}
\end{equation}

Note that $C_{\alpha}(p\Vert q)=(1-\alpha)R_{\alpha}(p\Vert q)$,
where $R_{\alpha}$ is \emph{R{\'e}nyi divergence of order $\alpha$
}\cite{van2014renyi}. 

We show that for any $\alpha\in\left[0,1\right]$, $\hat{H}_{C_{\alpha}}(X)$
is a lower bound on the entropy (for $\alpha\notin\left[0,1\right]$,
$C_{\alpha}$ is not a valid distance function (see Appendix \ref{appendix:chernoff-alpha-bounds})).
To do so, we make use of a derivation from~\cite{haussler_mutual_1997} and write,
\begin{align}
&H(X) \nonumber \\
&= H(X\vert C)-\int\sum_{i}c_{i}p_{i}(x)\ln\frac{p_{X}(x)}{p_{i}(x)}\,dx\nonumber \\
 & =H(X\vert C)-\int\sum_{i}c_{i}p_{i}(x)\ln\frac{p_{X}(x)}{{p_{i}\left(x\right)}^{\alpha}\sum_{j}c_{j}{p_{j}(x)}^{1-\alpha}}\,dx \nonumber \\
 &\quad -\int\sum_{i}c_{i}p_{i}(x)\ln\left(\frac{\sum_{j}c_{j}{p_{j}(x)}^{1-\alpha}}{{p_{i}(x)}^{1-\alpha}}\right)\,dx\nonumber \\
 & \stackrel{(a)}{\ge} H(X\vert C) -\int\sum_{i}c_{i}p_{i}(x)\ln\left(\frac{\sum_{j}c_{j}{p_{j}(x)}^{1-\alpha}}{{p_{i}(x)}^{1-\alpha}}\right)\,dx\nonumber \\
 & \stackrel{(b)}{\ge} H(X\vert C) -\sum_{i}c_{i}\ln\int{p_{i}(x)}^{\alpha}\sum_{j}c_{j}{p_{j}(x)}^{1-\alpha}\,dx\nonumber \\
 & =H(X\vert C)-\sum_{i}c_{i}\ln\sum_{j}c_{j}e^{-C_{\alpha}\left(p_{i}\Vert p_{j}\right)} := \hat{H}_{C_{\alpha}}(X)\,.\label{eq:chernoff_lower}
\end{align}

The inequalities (\emph{a}) and (\emph{b}) follow from Jensen's inequality. This
inequality is used directly in $(b)$, while in $(a)$ it follows
from
\begin{align*}
&-\int\sum_{i}c_{i}p_{i}(x)\ln\frac{p_{X}(x)}{{p_{i}\left(x\right)}^{\alpha}\sum_{j}c_{j}{p_{j}(x)}^{1-\alpha}}\,dx \\
&\ge-\ln\int\sum_{i}c_{i}p_{i}(x)\frac{p_{X}(x)}{{p_{i}\left(x\right)}^{\alpha}\sum_{j}c_{j}{p_{j}(x)}^{1-\alpha}}\,dx\\
 & =-\ln\int\sum_{i}c_{i}{p_{i}(x)}^{1-\alpha}\frac{p_{X}(x)}{\sum_{j}c_{j}{p_{j}(x)}^{1-\alpha}}\,dx\\
 & =-\ln \int p_{X}(x)\,dx=0.
\end{align*}

Note that Jensen's inequality is used in the derivations of both this lower bound as well as the lower bound $\hat{H}_{\text{ELK}}$ in Equation (\ref{eq:lowerbound-ELK}).  However, the inequality is applied differently in the two cases, and, as will be demonstrated in Section \ref{sec:Numerical-results}, the estimators have different performance.

We have shown that using $C_{\alpha}$ as a distance function gives a
lower bound on the mixture entropy for any $\alpha\in[0,1]$. For
a general mixture distribution, one could optimize over the value
of $\alpha$ to find the tightest lower bound. However,
we can show that the tightest bound is achieved for $\alpha=0.5$
 in the special case when all of the mixture components $p_{i}$ are symmetric and come from
a location family,
\[
p_{i}(x)=b(x-\mu_{i})=b(\mu_{i}-x) \,.
\]
Examples of this situation include mixtures of Gaussians with the
same covariance (``homoscedastic'' mixtures), multivariate $t$-distributions
with the same covariance, location-shifted bounded uniform distributions,
most kernels used in kernel density estimation, etc.  It does not apply to skewed distributions, such as 
as the skew-normal distribution~\cite{contreras2016bounds}.

To show that $\alpha=0.5$ is optimal, first define the Chernoff $\alpha$-coefficient
as 
\[
c_{\alpha}(p_{i}\Vert p_{j}):=\int{p_{i}(x)}^{\alpha}{p_{j}(x)}^{1-\alpha}dx\,.
\]

We show that for any pair $p_{i},p_{j}$ of symmetric distributions from a location family,
$c_{\alpha}(p_{i}\Vert p_{j})$ is minimized by $\alpha=0.5$. This
means that all pairwise distances $C_{\alpha}(p_i\Vert p_j) \equiv -\ln c_\alpha(p_i\Vert p_j)$ are \emph{maximized
}by $\alpha=0.5$, and, therefore, the entropy estimator $H_{C_{\alpha}}$
(Equation (\ref{eq:pairwise})) is maximized by $\alpha=0.5$.

First, define a change of variables
\[
y:=\mu_{i}+\mu_{j}-x\,,
\]
which gives $x-\mu_{i}=\mu_{j}-y$ and $x-\mu_{j}=\mu_{i}-y$. This
allows us to write the Chernoff $\alpha$-coefficient as
\begin{align*}
c_{\alpha}(p_{i}\Vert p_{j})= & \int{p_{i}(x)}^{\alpha}{p_{j}(x)}^{1-\alpha}dx\\
= & \int{b(x-\mu_{i})}^{\alpha}{b(x-\mu_{j})}^{1-\alpha}dx\\
\stackrel{(a)}{=} & \int{b(\mu_{j}-y)}^{\alpha}{b(\mu_{i}-y)}^{1-\alpha}dy\\
\stackrel{(b)}{=} & \int{b(y-\mu_{j})}^{\alpha}{b(y-\mu_{i})}^{1-\alpha}dy\\
= & c_{1-\alpha}(p_{i}\Vert p_{j}),
\end{align*}
where, in $(a)$, we have substituted variables, and in $(b)$ we used
the assumption that $b(x)=b(-x)$. Since we have shown that 
$c_{\alpha}(p_{i}\Vert p_{j})=c_{1-\alpha}(p_{i}\Vert p_{j})$,
$c_{\alpha}$ is symmetric in $\alpha$ about $\alpha=0.5$. In Appendix \ref{appendix:chernoff-alpha-bounds}, we show that $c_{\alpha}(p\Vert q)$ is everywhere convex
in $\alpha$. Together, this means that $c_{\alpha}(p_{i}\Vert p_{j})$
must achieve a minimum value at {$\alpha=0.5$}. 

The Chernoff $\alpha$-coefficient for $\alpha=0.5$ is known as
the \emph{Bhattacharyya coefficient}, with the corresponding \emph{Bhattacharyya
distance}~\cite{fukunaga_introduction_1990} defined as
\begin{align*}
\text{BD}(p\Vert q) & :=-\ln\int\sqrt{p(x)q(x)}dx=C_{0.5}(p\Vert q).
\end{align*}

Since any Chernoff $\alpha$-divergence is a lower bound for the entropy,
we write the particular case of Bhattacharyya-distance lower bound
as
\begin{equation}
H(X)\ge\hat{H}_{\text{BD}}(X):=H(X\vert C)-\sum_{i}c_{i}\ln\sum_{j}c_{j}e^{-\text{BD}(p_{i}\Vert p_{j})}\,.\label{eq:bhat-lower}
\end{equation}
\vspace{-18pt}
\subsection{Upper Bound\label{subsec:Upper-bound}}

The \emph{Kullback--Leibler {[}KL{]} divergence}~\cite{cover_elements_2012_newversion} is
defined as
\[
\text{KL}(p\Vert q):=\int p(x)\ln\frac{p(x)}{q(x)}dx\,.
\]
Using KL divergence as the pairwise distance provides an upper bound on the mixture entropy. We show
this as follows:
\begin{align*}
H(X)& = -\sum_{i}c_{i}\text{E}_{p_{i}}\left[\ln\sum_{j}c_{j}p_{j}(X)\right]\\
 & \stackrel{(a)}{\le} -\sum_{i}c_{i}\ln\sum_{j}c_{j}e^{\text{E}_{p_{i}}[\ln p_{j}(X)]}\\
 & = -\sum_{i}c_{i}\ln\sum_{j}c_{j}e^{-H(p_{i}\Vert p_{j})}\\
 & =  \sum_{i}c_{i}H(p_{i})-\sum_{i}c_{i}\ln\sum_{j}c_{j}e^{-\text{KL}(p_{i}\Vert p_{j})},
\end{align*}
where $\text{E}_{p_i}$ indicates expectation when $X$ is distributed according to $p_i$, $H(\cdot\Vert\cdot)$ indicates the cross-entropy function,
and we employ the identity $H(p_{i}\Vert p_{j})=H(p_{i})+\text{KL}(p_{i}\Vert p_{j})$.
The inequality in step $(a)$ uses a variational lower bound on the
expectation of a log-sum~\cite{paisley2010two,hershey_approximating_2007},
\[
\text{E}\left[\ln\sum_{j}Z_{j}\right]\ge\ln\sum_{j}e^{\text{E}\left[\ln Z_{j}\right]}\,.
\]

Combining yields the upper bound
\begin{equation}
H(X)\le\hat{H}_{\text{KL}}:=H(X\vert C)-\sum_{i}c_{i}\ln\sum_{j}c_{j}e^{-\text{KL}(p_{i}\Vert p_{j})}\,.\label{eq:kl-upper-bound}
\end{equation}

\subsection{Exact Estimation in the ``Clustered'' Case\label{subsec:clustered}}

In the previous sections, we derived lower and upper bounds on
the mixture entropy, using estimators based on Chernoff $\alpha$-divergence
and KL divergence, respectively.

There are situations in which the lower and upper bounds become similar.  Consider a pair of component distributions, $p_i$ and $p_j$. By applying Jensen's inequality to Equation (\ref{eq:chernoff-divergence}), we can derive the inequality $C_\alpha(p_i \Vert p_j) \le \alpha \text{KL}(p_i \Vert p_j)$.  There are two cases in which a pair of components contributes similarly to the lower and upper bounds. The first case is when $C_\alpha(p_i \Vert p_j)$ is very large, meaning that the KL is also very large. By Equation (\ref{eq:pairwise}), distances enter into our estimators as $e^{-D(p_i\Vert p_j)}$, and, in this case, $e^{-\text{KL}(p_i\Vert p_j)}  \approx e^{-C_\alpha(p_i\Vert p_j)} \approx 0$.   In the second case, $\text{KL}(p_i \Vert p_j) \approx 0$, meaning that $C_\alpha(p_i \Vert p_j)$ must also be near zero, and, in this case, $e^{-\text{KL}(p_i \Vert p_j)} \approx e^{-C_\alpha(p_i\Vert p_j)} \approx 1$. Thus, 
the lower and upper bounds become similar when all pairs of components are either very close together or very far apart.  

In this section, we analyze this special case.
Specifically, we consider the situation when mixture components are ``clustered'',
meaning that there is a grouping of component distributions
such that distributions in the same group are approximately the same and
distributions assigned to different groups are very different
from one another.
We show that in this case our lower and upper bounds become equal and our pairwise-distance estimate of the entropy is tight.  Though this situation may seem like an edge case,
clustered distributions do arise in mixture estimation, e.g., when
there are repeated data points, or as solutions
to information-theoretic optimization problems~\cite{kolchinsky_nonlinear_2017}.
Note that the number of groups is arbitrary, and therefore this situation
includes the extreme cases of a single group (all component distributions
are nearly the same) as well as $N$ different groups (all component
distributions are very different).

Formally, let the function $g(i)$ indicate the group of component
$i$. We define that the components are ``clustered'' with respect
to grouping $g$ iff $\text{KL}(p_{i}\Vert p_{j})\le\kappa$ whenever
$g(i)=g(j)$ for some small $\kappa$, and $\text{BD}(p_{i}\Vert p_{j})\ge\beta$
whenever $g(i)\ne g(j)$ some large $\beta$. We use the notation
$p_{G}(k)=\sum_{i}\delta_{g(i),k}c_{i}$ to indicate the sum of the
weights of the components in group $k$, where $\delta_{ij}$ indicates the Kronecker delta function.
For technical reasons, below we only consider $C_\alpha$ where $\alpha$ is strictly greater than 0.

We show that when $\kappa$ is small and $\beta$ is large, both $\hat{H}_{C_{\alpha}}$
for $\alpha\in(0,1]$ and $\hat{H}_{\text{KL}}$ approach
\[
H(X\vert C)-\sum_{k}p_{G}(k)\ln p_{G}(k)\,.
\]

Since one is a lower bound and one is an upper bound on the true entropy,
the estimators become exact as they converge in value.

Recall that $\text{BD}(p\Vert q)=C_{0.5}(p\Vert q)$. For $\alpha\in(0,1]$, $\alpha^{-1}C_{\alpha}(p\Vert q)$
is a monotonically decreasing function for $\alpha\in(0,1]$~\cite{sason2016f},
meaning that $C_{\alpha}\ge2\alpha\text{BD}(p\Vert q)$ for $\alpha\in(0,0.5]$.
In addition, $(1-\alpha)^{-1}C_{\alpha}(p\Vert q)$ is a monotonically
increasing function for $\alpha>0$~\cite{sason2016f}, thus $C_{\alpha}\ge2(1-\alpha)\text{BD}(p\Vert q)$
for $\alpha\in[0.5,1]$. Using the assumption that $\text{BD}(p_{i}\Vert p_{j})\ge\beta$
and combining gives the bound 
\[
C_{\alpha}(p\Vert q)\ge(1-\left|1-2\alpha\right|)\beta
\]
for $\alpha\in(0,1]$, leading to
\begin{align*}
\hat{H}_{C_{\alpha}}(X) & :=H(X\vert C)-\sum_{i}c_{i}\ln\sum_{j}c_{j}e^{-C_{\alpha}(p_{i}\Vert p_{j})}\\
 & \ge H(X\vert C)-\sum_{i}c_{i}\ln\Big[\sum_{j}\delta_{g(i),g(j)}c_{j}+ \\
 & \qquad\quad \sum_{j}\left(1-\delta_{g(i),g(j)}\right)c_{j}e^{-C_{\alpha}(p_{i}\Vert p_{j})}\Big]\\
 & \ge H(X\vert C)-\sum_{k}p_{G}(k)\ln\Big[p_{G}(k)+\\
 &\qquad\quad (1-p_{G}(k))e^{-(1-\left|1-2\alpha\right|)\beta}\Big]\,.
\end{align*}

In the second line, for the summation over $i,j$ in the same group, we've used the non-negativity of
$C_{\alpha}(p_{i}\Vert p_{j})\ge0$.

For the upper bound $\hat{H}_{\text{KL}}$, we use that $\text{KL}(p_{i}\Vert p_{j})\le\kappa$
for $i$ and $j$ in the same group, and otherwise $e^{-\text{KL}(p_{i}\Vert p_{j})}\ge0$.
This gives the bound
\begin{align*}
\hat{H}_{\text{KL}}(X) & :=H(X\vert C)-\sum_{i}c_{i}\ln\sum_{j}c_{j}e^{-\text{KL}(p_{i}\Vert p_{j})}\\
 & \le H(X\vert C)-\sum_{i}c_{i}\ln\Big[\sum_{j}\delta_{g(i),g(j)}c_{j}e^{-\kappa}+\\
 &\qquad \quad \sum_{j}\left(1-\delta_{g(i),g(j)}\right)c_{j}e^{-\text{KL}(p_{i}\Vert p_{j})}\Big]\\
 & \le H(X\vert C)-\sum_{k}p_{G}(k)\ln p_{G}(k)e^{-\kappa} \,.
\end{align*}

The difference between the bounds is bounded by
\begin{align*}
& \hat{H}_{\text{KL}}(X)-\hat{H}_{C_{\alpha}}(X) \\
& \le\kappa+\sum_{k}p_{G}(k)\ln\left[1+\frac{(1-p_{G}(k))e^{-(1-\left|1-2\alpha\right|)\beta}}{p_{G}(k)}\right]\\
 & \le\kappa+\sum_{k}p_{G}(k)\frac{(1-p_{G}(k))e^{-(1-\left|1-2\alpha\right|)\beta}}{p_{G}(k)}\\
 & =\kappa+\left(\left|G\right|-1\right)e^{-(1-\left|1-2\alpha\right|)\beta} \,,
\end{align*}
where $\left|G\right|$ is the number of groups. Thus, the difference
decreases at least linearly in $\kappa$ and exponentially in $\beta$.
This shows that, in the clustered case, when $\kappa\approx0$ and
$\beta$ is very large, our lower and upper bounds become exact.

It also shows that any distance measure bounded between BD and KL
also gives an exact estimate of entropy in the clustered case. Furthermore,
the idea behind this proof can be extended to estimators induced by
other bounding distances, beyond BD and KL, so as to show
that a particular estimator converges to an exact entropy estimate
in the clustered case. Note, however, that, for some distribution-distance
pairs, the components will never be considered as ``clustered'';
e.g., the $\alpha$-Chernoff distance for $\alpha=0$ between any
two Gaussians is $0$, and so a Gaussian mixture distribution will
never be considered clustered according to this distance.

Finally, in the perfectly clustered case, we can show that our lower bound, $\hat{H}_{\text{BD}}$, is at least as good as the Expected Likelihood Kernel lower bound, $\hat{H}_{\text{ELK}}$, as defined in Equation (\ref{eq:lowerbound-ELK}).  See Appendix \ref{appendix:bd-better-ekl} for details.

\section{Gaussian Mixtures\label{sec:Entropy-of-Gaussian}}

Gaussians are very frequently used as components in mixture distributions.
Our family of estimators is well-suited to estimating the entropies
of Gaussian mixtures, since the entropy of a $d$-dimensional Gaussian
$p_{i} = \mathcal{N}\left(\mu_{i},\bm{\Sigma}_{i}\right)$ has a
simple closed-form expression,
\begin{equation}
H\left(p_{i}\right)=\frac{1}{2}\left[\ln\left|\bm{\Sigma}_{i}\right|+d\ln2\pi+d\right]\,,\label{eq:gaussianent}
\end{equation}
and because there are many distance functions between Gaussians with
closed-form expressions (KL divergence, the Chernoff $\alpha$-divergences~\cite{hero_alpha-divergence_2001},
2-Wasserstein distance~\cite{dowson1982frechet,olkin1982distance}, etc.). In this section, we consider Gaussian
mixtures and state explicit expression for the lower and upper bounds
on the mixture entropy
derived in the previous section. We also consider these bounds in
the special case where all Gaussian components have the same covariance
matrix (homoscedastic mixtures).

We first consider the lower bound, $\hat{H}_{C_{\alpha}}$, based
on the Chernoff $\alpha$-divergence distance function. For two multivariate
Gaussians $p_{1} = \mathcal{N}\left(\mu_{1},\bm{\Sigma}_{1}\right)$
and $p_{2} = \mathcal{N}\left(\mu_{2},\bm{\Sigma}_{2}\right)$, this
distance is defined as~\cite{hero_alpha-divergence_2001}:
\begin{multline}
C_{\alpha}(p_{1}\Vert p_{2})=\\
\frac{\left(1-\alpha\right)\alpha}{2}(\mu_{1}-\mu_{2})^{T}\left((1-\alpha)\bm{\Sigma}_{1}+\alpha\bm{\Sigma}_{2}\right)^{-1}(\mu_{1}-\mu_{2})+\\
\frac{1}{2}\ln\left(\frac{\left|(1-\alpha)\bm{\Sigma}_{1}+\alpha\bm{\Sigma}_{2}\right|}{\left|\bm{\Sigma}_{1}\right|^{1-\alpha}\left|\bm{\Sigma}_{2}\right|^{\alpha}}\right)
\,.
\label{eq:chernoff-normals}
\end{multline}

(As a warning, note that most sources show erroneous expressions for
the Chernoff and/or R{\'e}nyi $\alpha$-divergence between two multivariate
Gaussians, including~\cite{gil2013renyi, nielsen_chernoff_2011, pardo_statistical_2005, hobza_renyi_2009, nielsen_generalized_2014},
and even a late draft of this manuscript.)

For the upper bound $\hat{H}_{\text{KL}}$, the KL divergence between two
multivariate Gaussians $p_{1}\sim\mathcal{N}\left(\mu_{1},\bm{\Sigma}_{1}\right)$
and $p_{2}\sim\mathcal{N}\left(\mu_{2},\bm{\Sigma}_{2}\right)$ is
\begin{multline}
\text{KL}(p_{1}\Vert p_{2}) =  \frac{1}{2}\Big[\ln\left|\bm{\Sigma}_{2}\right|-\ln\left|\bm{\Sigma}_{1}\right|+\\
\left(\mu_{1}-\mu_{2}\right)^{T}\bm{\Sigma}_{2}^{-1}\left(\mu_{1}-\mu_{2}\right)+\tr\left(\bm{\Sigma}_{2}^{-1}\bm{\Sigma}_{1}\right)-d\Big]\,.\label{eq:gaussian-kl}
\end{multline}

The appropriate lower and upper bounds are found by plugging
in Equations (\ref{eq:chernoff-normals}) and (\ref{eq:gaussian-kl}) into Equation (\ref{eq:pairwise}).

These bounds have simple forms when all of the mixture components
have equal covariance matrices, i.e., $\bm{\Sigma}_{i}=\bm{\Sigma}$
for all $i$. First, define a transformation in which each Gaussian
component $p_{j}$ is mapped to a different Gaussian $\tilde{p}_{j,\alpha}$,
which has the same mean but where the covariance matrix is rescaled
by $\frac{1}{\alpha(1-\alpha)}$,
\[
p_{j}:=\mathcal{N}\left(\mu_{j},\bm{\Sigma}\right)\quad\mapsto\quad\tilde{p}_{j,\alpha}:=\mathcal{N}\left(\mu_{j},\frac{1}{\alpha(1-\alpha)}\bm{\Sigma}\right)\,.
\]

Then, the lower bound of Equation (\ref{eq:chernoff_lower}) can be written as
\[
\hat{H}_{C_{\alpha}}=\frac{d}{2}+\frac{d}{2}\ln\left(\alpha(1-\alpha)\right)-\sum_{i}c_{i}\ln\sum_{j}c_{j}\tilde{p}_{j,\alpha}(\mu_{i})\,.
\]
This is derived by combining the expressions for $C_{\alpha}$, Equation (\ref{eq:chernoff-normals}),
the entropy of a Gaussian, Equation (\ref{eq:gaussianent}), and the Gaussian density
function. For a homoscedastic mixture, the tightest lower bound among
the Chernoff $\alpha$-divergences is given by $\alpha=0.5$, corresponding
to the Bhattacharyya distance,
\[
\hat{H}_{\text{BD}}=\frac{d}{2}+\frac{d}{2}\ln\frac{1}{4}-\sum_{i}c_{i}\ln\sum_{j}c_{j}\tilde{p}_{j,0.5}(\mu_{i})\,.
\]
(This is derived above in Section \ref{subsec:Lower-Bound}.)

{For the upper bound, when all Gaussians have the same covariance matrix,
we again combine the expressions for $\text{KL}$, Equation (\ref{eq:gaussian-kl}),
the entropy of a Gaussian, Equation (\ref{eq:gaussianent}), and the Gaussian density function to give}
\[
\hat{H}_{\text{KL}}(X)=\frac{d}{2}-\sum_{i}c_{i}\ln\sum_{j}c_{j}p_{j}(\mu_{i})\,.
\]

Note that this is exactly the expression for the kernel density estimator
$\hat{H}_{\text{KDE}}$ (Equation (\ref{eq:h-kde})), plus a dimensional correction.
Thus, surprisingly $\hat{H}_{\text{KDE}}$ is a reasonable entropy
estimator for homoscedastic Gaussian mixtures, since it is only an
additive constant away from KL-distance based estimator $\hat{H}_{\text{KL}}$
(which has various beneficial properties, as described above). This
may explain why $\hat{H}_{\text{KDE}}$ has been used effectively
in optimization contexts~\cite{schraudolph_optimization_1995,viola_empirical_1996,schraudolph_gradient-based_2004,shwartz_fast_2005},
where the additive constant is often irrelevant, despite lacking a
principled justification in terms of being a a bound on entropy.

\section{Estimating Mutual Information\label{sec:Mutual-information-across}}

It is often of interest, for example in rate distortion problems and
related problems~\cite{kolchinsky_nonlinear_2017}, to calculate the
mutual information across a communication channel,
\[
MI(X;U)=H(X)-H(X\vert U),
\]
where $U$ is the distribution of signals sent across the channel,
and $X$ is the distribution of signals received on the other end
of the channel. As with mixture distributions, it is often easy
to compute $H(X\vert U)$, the entropy of the received signal given
the sent signal (i.e., the distribution of noise on the channel).
The marginal entropy of the received signals, $H(X)$, on the other hand, is
often difficult to compute.

In some cases, the distribution over sent signals $U$ may be well approximated by
mixture of $N$ components,
\[
p_{U}(u)=\sum_{i=1}^{N} c_{i} p_{i}(u)\,.
\]
In that case, the distribution over received signals $X$ can also be written as a mixture of $N$ components,
\begin{multline*}
p_{X}(x) = \int p_{X\vert U}(x\vert u) \left(  \sum_{i=1}^{N} c_{i} p_{i}(u) \right) \; du  \\
= \sum_{i=1}^{N} c_i \left( \int p_{X\vert U}(x\vert u) p_{i}(u) \; du \right) =  \sum_{i=1}^{N} c_i q_i(x)  \,,
\end{multline*}
where $p_{X\vert U}$ is the conditional probability distribution of the channel, and where we have defined $q_i(x) := \int p_{X\vert U}(x\vert u) p_{i}(u) \; du$ as the $i^\text{th}$ component of the mixture distribution over $X$.

In situations where the components $q_i$  have closed-form expressions for entropy and divergence (e.g., when $p_i$ are Gaussians and the channel noise is Gaussian and homoscedastic, the $q_i$ will themselves be Gaussians, as discussed below),  %
we can provide lower and upper bounds on the mutual information $MI(X;U)$ by bounding $H(X)$ using our pairwise distance estimators. %
In particular, we
have the lower bound
\begin{multline}
MI(X;U)  = H(X) - H(X\vert U) \\
\qquad \ge  - \sum_i c_i \ln \sum_j c_j e^{-C_\alpha(q_i \Vert q_j)} + H(X \vert C) - H(X\vert U) \,,
\end{multline}
and the upper bound,
\begin{multline}
MI(X;U)  = H(X) - H(X\vert U) \\
\qquad \le  - \sum_i c_i \ln \sum_j c_j e^{-\text{KL}(q_i \Vert q_j)} + H(X \vert C) - H(X\vert U) \,.
\end{multline}
We remind the readers that $H(X \vert C)=\sum_i c_i H(q_i)$ represents the average entropy of the components of $X$, while $H(X\vert U)$ represents the conditional entropy of $p_{X\vert U}$, the channel from $U$ to $X$.  

As a practical example, consider a scenario in which $U$ is a random
variable representing outside temperature on any particular day. This
temperature is measured with a thermometer with homoscedastic Gaussian measurement
noise (the ``Additive White Noise Gaussian channel''). This gives
our measurement distribution
\[
X=U+\mathcal{N}\left(0,\bm{\Sigma}'\right).
\]
If the actual temperature distribution is 
distributed as a mixture of $N$ Gaussians, each one having mixture
weight $c_{i}$, mean $\mu_{i}$, and covariance matrix $\bm{\Sigma_{i}}$,
then $X$ will also be distributed as a mixture of $N$ Gaussians,
each with weight $c_{i}$, mean $\mu_{i}$, and covariance matrix
$\bm{\Sigma}_{i}^{\prime}:=\bm{\Sigma}_{i}+\bm{\Sigma}'$. Combining our estimators of the entropy of the mixture $X$ with closed-form expressions for
$H(X\vert C)$ (the weighted average of the entropies of $\{ \mathcal{N}(\mu_i, \bm{\Sigma}_{i}+\bm{\Sigma}')\}_{i=1..N}$) and $H(X\vert U)$ (the entropy of $\mathcal{N}(0, \bm{\Sigma}')$) gives lower and upper bounds on the  mutual information between actual temperature, $U$, and thermometer measurements, $X$.

\section{Numerical Results\label{sec:Numerical-results}}

In this section, we run numerical experiments and compare estimators
of mixture entropy under a variety of conditions. We consider two
different types of mixtures, mixtures of Gaussians and mixtures of
uniform distributions, for a variety of parameter values. We evaluate
the following estimators:
\begin{enumerate} %
\item The true entropy, $H(X)$, as estimated by a Monte Carlo sampling
of the mixture model. Two thousand samples were used for each MC estimate
for the mixtures of Gaussians, and 5000 samples were used for the
mixtures of uniform distributions.
\item Our proposed upper-bound, based on the KL divergence, $\hat{H}_{\text{KL}}$
(Equation (\ref{eq:kl-upper-bound}))
\item Our proposed lower-bound, based on the Bhattacharyya distance, $\hat{H}_{\text{BD}}$
(Equation (\ref{eq:bhat-lower}))
\item The kernel density estimate based on the component means, $\hat{H}_{\text{KDE}}$
(Equation (\ref{eq:h-kde}))
\item The lower bound based on the ``Expected Likelihood Kernel'', $\hat{H}_{\text{ELK}}$
(Equation (\ref{eq:lowerbound-ELK}))
\item The lower bound based on the conditional entropy, $H(X\vert C)$ (Equation (\ref{eq:conditional}))
\item The upper bound based on the joint entropy, $H(X,C)$ (Equation (\ref{eq:joint})).
\end{enumerate}

We show the values of the estimators 1--5 as line plots, while the
region between the conditional (6) and joint entropy (7) is shown
in shaded green. The code for these figures can be found at~\cite{codeurl},
and uses the Gonum numeric library~\cite{gonumurl}.

\subsection{Mixture of Gaussians}

\begin{figure*}
\begin{subfigure}[H]{0.49\linewidth}
\subplotlabel{\includegraphics[width=\figwidth\linewidth]{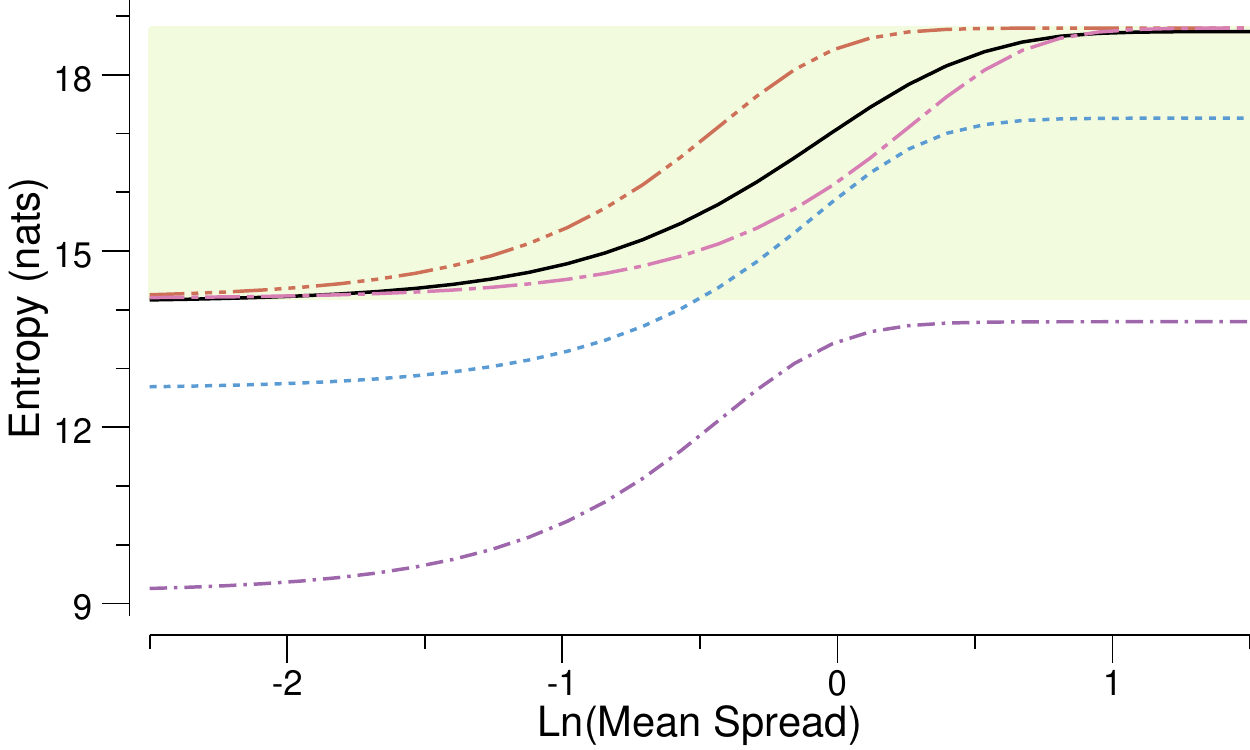}}{A}
\vspace{15pt}
\end{subfigure}
\begin{subfigure}[H]{0.49\linewidth}
\subplotlabel{

\includegraphics[width=\figwidth\linewidth]{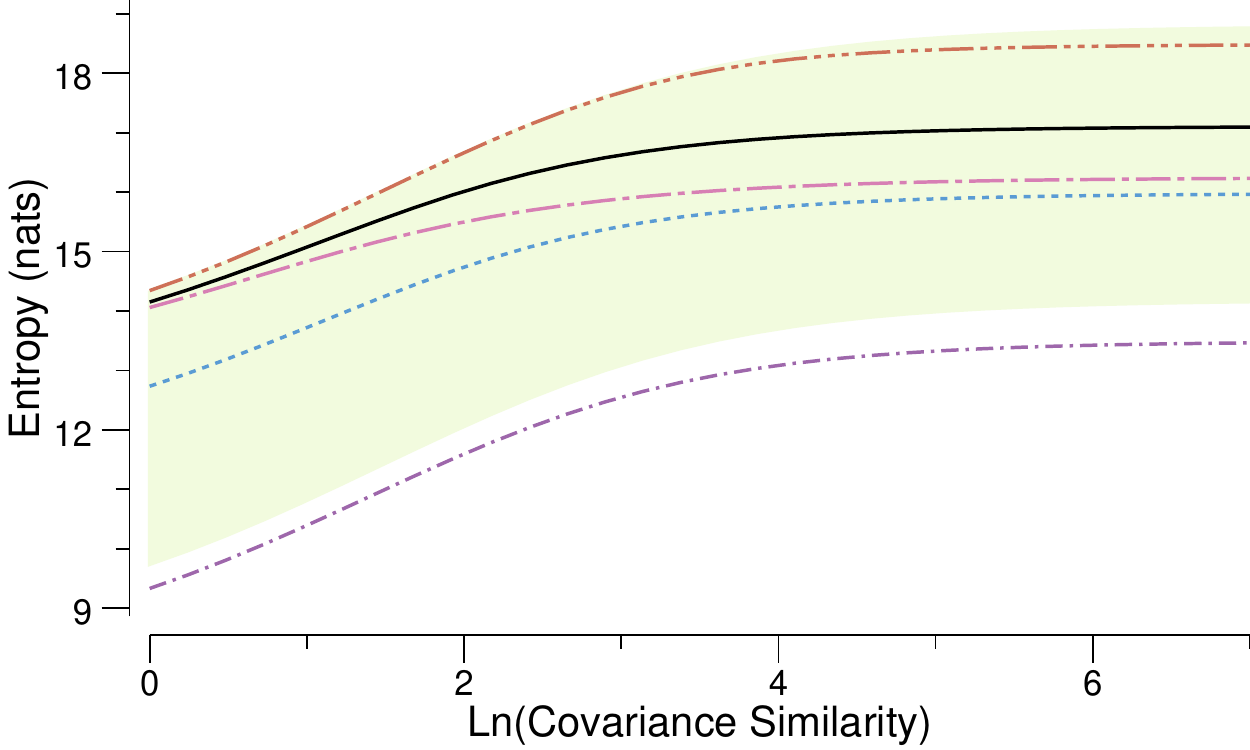}

}{B}\vspace{15pt}
\end{subfigure}
\\
\begin{subfigure}[H]{0.49\linewidth}
\subplotlabel{

\includegraphics[width=\figwidth\linewidth]{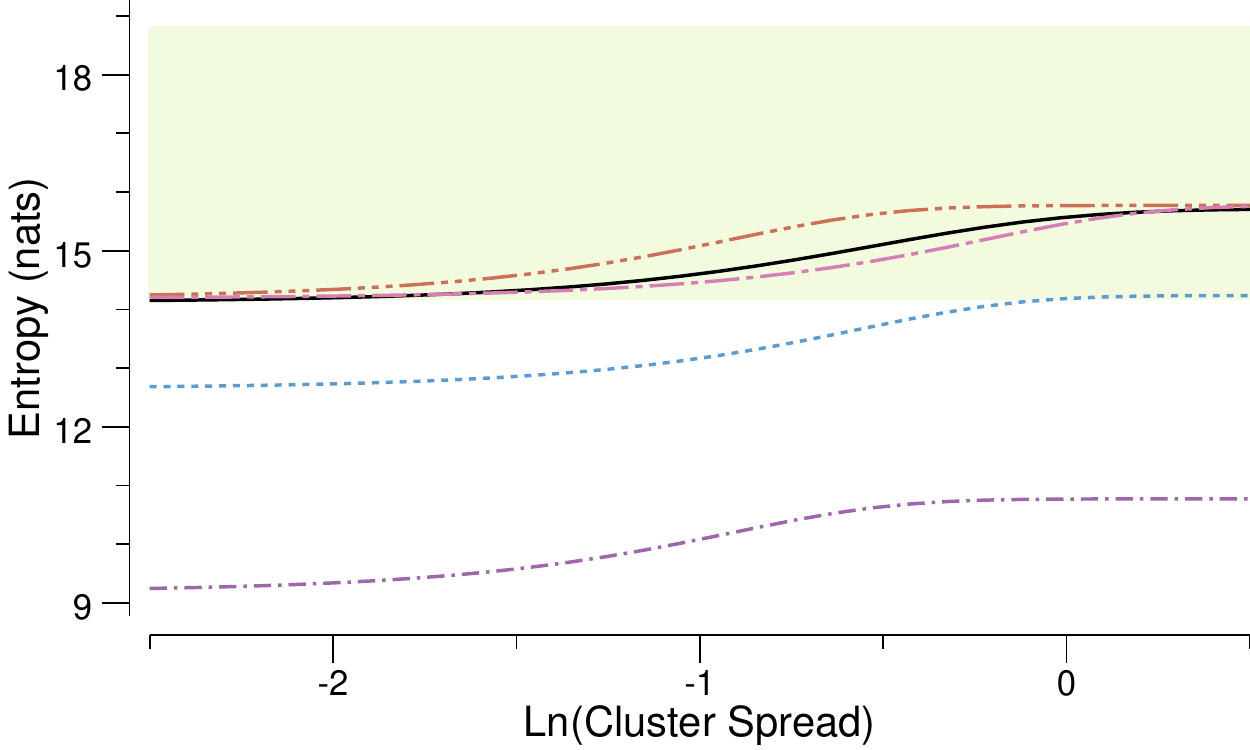}

}{C}
\end{subfigure}
\begin{subfigure}[H]{0.49\linewidth}
\subplotlabel{
\begin{raggedright}
\includegraphics[width=\figwidth\linewidth]{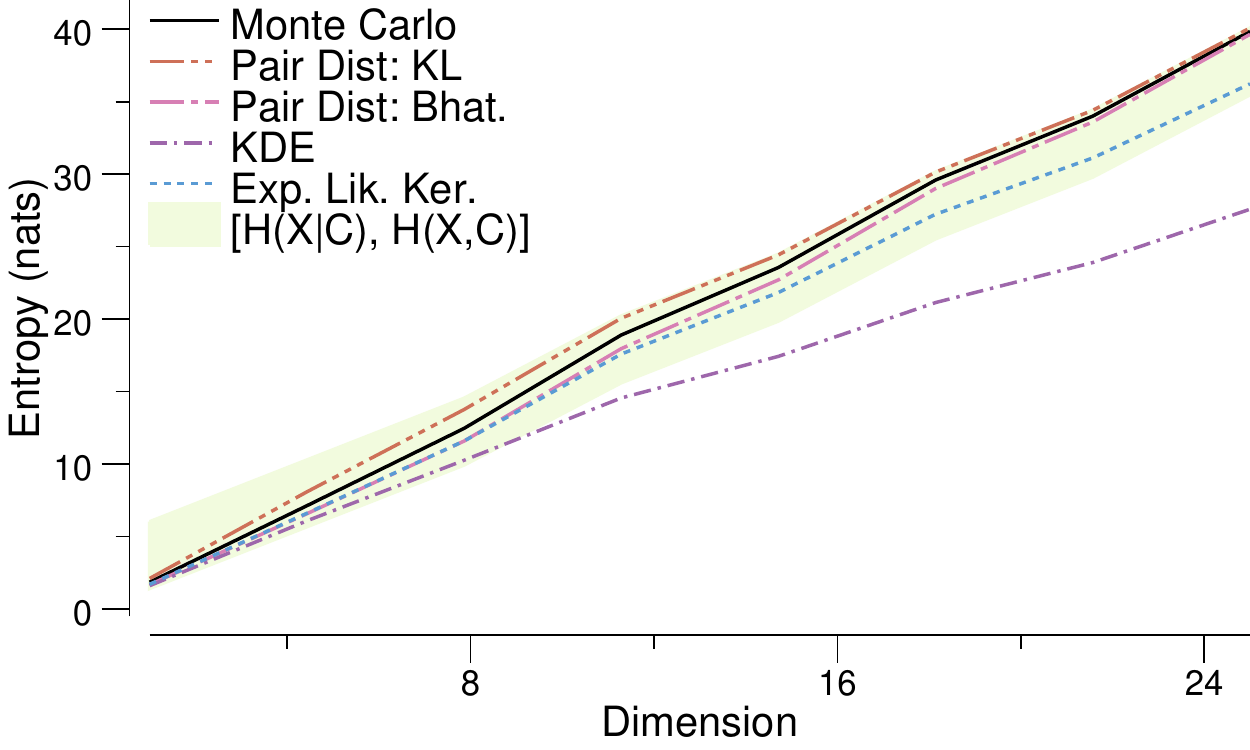}
\par\end{raggedright}
}{D}
\end{subfigure}

\caption{Entropy estimates for a mixture of a 100 Gaussians. In each
plot, the vertical axis shows the entropy of the distribution, and
the horizontal axis changes a feature of the components: (\textbf{A}) the distance between means is increased; (\textbf{B})
the component covariances become more similar (at the right side of
the plot, all Gaussians have covariance matrices approximately equal to the identity matrix;
(\textbf{C}) the components are grouped into five ``clusters'', and the distance
between the locations of the clusters is increased;
(\textbf{D})
the dimension is increased.\label{fig:centerspread}}
\end{figure*}

In the first experiment, we evaluate the estimators on a mixture of
randomly placed Gaussians, and look at their behavior as the distance
between the means of the Gaussians increases. The mixture is composed
of 100 10-dimensional Gaussians, each Gaussian distributed as 
$p_{i} = \mathcal{N}\left(\mu_{i},\bm{I}_{(10)}\right)$,
where $\bm{I}_{(d)}$ indicates the $d \times d$ identity matrix.  
Means are sampled from $\mu_{i}\sim \mathcal{{N}}(0,\sigma\bm{I}_{(10)})$.
Figure \ref{fig:centerspread}A depicts the change in estimated entropy as
the means grow farther apart, in particular a function of $\ln(\sigma)$.
We see that
 our proposed bounds are closer to the true entropy than the other estimators
 over the whole range of $\sigma$ values, and in the extremes, our
bounds approach the exact value of the true entropy. This is as expected,
since as $\sigma\rightarrow0$ all of the Gaussian mixture components
become identical, and as $\sigma\rightarrow\infty$ all of the Gaussian
components grow very far apart, approaching the case where each Gaussian
is in its own ``cluster''. The ELK lower bound is a strictly worse
estimate than $\hat{{H}}_{\text{BD}}$, in this experiment. As expected,
the KDE estimator differs by exactly $d/2$ from
the KL estimator. 

In the second experiment, we evaluate the entropy estimators as the
covariance matrices change from less to more similar. We again generate
100 10-dimensional Gaussians. Each Gaussian is distributed as $p_{i} = \mathcal{N}\left(\mu_{i},\bm{\Sigma}_{i}\right)$,
where now $\mu_{i}\sim \mathcal{{N}}(0,\bm{I}_{(10)})$ and $\bm{\Sigma}_{i}\sim\mathcal{{W}}(\frac{1}{10+n}\bm{I}_{(10)},n)$,
where $\mathcal{{W}}(\bm{V}, n)$ is a Wishart distribution with scale-matrix $\bm{V}$ and $n$ degrees
of freedom. Figure \ref{fig:centerspread}B compares the the estimators with
the true entropy as a function of $\ln(n)$. When $n$ is small, the
Wishart distribution is broad and the covariance matrices differ significantly
from one another, while as $n\rightarrow\infty$, all the covariance
matrices become close to the identity $\bm{I}_{(10)}$. Thus, for small $n$,
we essentially recover a ``clustered'' case, in which every component
is in its own cluster and our lower and upper bounds give highly
accurate estimates. For large $n$, we converge to the $\sigma=1$
case of the first experiment.

In the third experiment, we again generate a mixture of 100 10-dimensional
Gaussians. Now, however, the Gaussians are grouped into five ``clusters'',
with each Gaussian component randomly assigned to one of the clusters.
We use $g(i)\in\left\{ 1\dots 5\right\} $ to indicate the group of each
Gaussian's component $i\in\left\{ 1 \dots 100\right\} $, and each of the
100 Gaussians is distributed as $p_{i} = \mathcal{N}(\tilde{\mu}_{g(i)},\bm{I}_{(10)})$.
The cluster centers $\tilde{\mu}_{k}$ for $k\in\left\{ 1 \dots 5\right\}$
are drawn from $\mathcal{{N}}(0,\sigma\bm{I}_{(10)})$. The results are depicted
in Figure~\ref{fig:centerspread}C as a function of $\ln(\sigma)$. 
In the first experiment, we saw that
the joint entropy $H(X,C)$ became an increasingly better estimator
as the Gaussians grew increasingly far apart. Here, however, we see
that there is a significant difference between $H(X,C)$ and the true
entropy, even as the groups become increasingly separated. Our proposed
bounds, on the other hand, provide accurate estimates of the entropy
across the entire parameter sweep. As expected, they become exact
in the limit when all clusters are at the same location, as well as
when all clusters are very far apart from each other.

Finally, we evaluate the entropy estimators while changing the dimension
of the Gaussian components. We again generate 100 Gaussian components,
each distributed as $p_{i} = \mathcal{{N}}(\mu_{i},\bm{I}_{(d)})$, with $\mu_{i}\sim \mathcal{{N}}(0,\sigma\bm{I}_{(d)})$.
We vary the dimensionality $d$ from $1$ to $60$. The results
are shown in Figure \ref{fig:centerspread}D. First, we see that when $d=1$,
the KDE estimator and the KL-divergence based estimator give a very
similar prediction (differing only by $0.5$), but as the dimension
increases, the two estimates diverge at a rate of $d/2$. Similarly,
$\hat{{H}}_{\text{ELK}}$ grows increasingly less accurate as the
dimension increases. Our proposed lower and upper bounds provide good
estimates of the mixture entropy across the whole sweep across dimensions.

As previously mentioned, our lower and upper bounds tend to perform best at
the ``extremes'' and worse in the intermediate 
regimes.
In particular, in Figures~\ref{fig:centerspread}A,C,D, the distances between component
means increase from left to right. 
On the left hand side of these figures, all of the component means are close and the 
component distributions overlap, as
evidenced by the fact that the mixture entropy is  $\approx H(X|C)$, i.e., $I(X;C)\approx 0$.  In this regime, when there is essentially a single ``cluster'', and our bounds become tight (see Section \ref{subsec:clustered}).
On the right hand side of these figures, the components' means are all far apart from each other,
and the mixture entropy  $\approx H(X,C)$, i.e., $I(X;C)\approx H(C)$
(in Figure~\ref{fig:centerspread}C, it is the five clusters that become far apart, and the mixture entropy $\approx H(X|C) + \ln 5$).  In this regime where there are many well-separated clusters, our bounds again become tight.
In between these two extremes, however,
there is no clear clustering of the mixture components, 
and the entropy bounds are not as tight.

As noted in the previous paragraph, the extremes in three out of the four subfigures approach the perfectly clustered case.
In this situation, we show in Appendix \ref{appendix:bd-better-ekl} that the BD-based estimator is a better bound on the true entropy than the 
Expected Likelihood Kernel estimator. We see confirmation of this in the experimental results, where $\hat{H}_{\text{ELK}}$
performs worse than the pairwise-distance based estimators.

\begin{figure*}
\begin{subfigure}[H]{0.49\linewidth}
\subplotlabel{\includegraphics[width=\figwidth\linewidth]{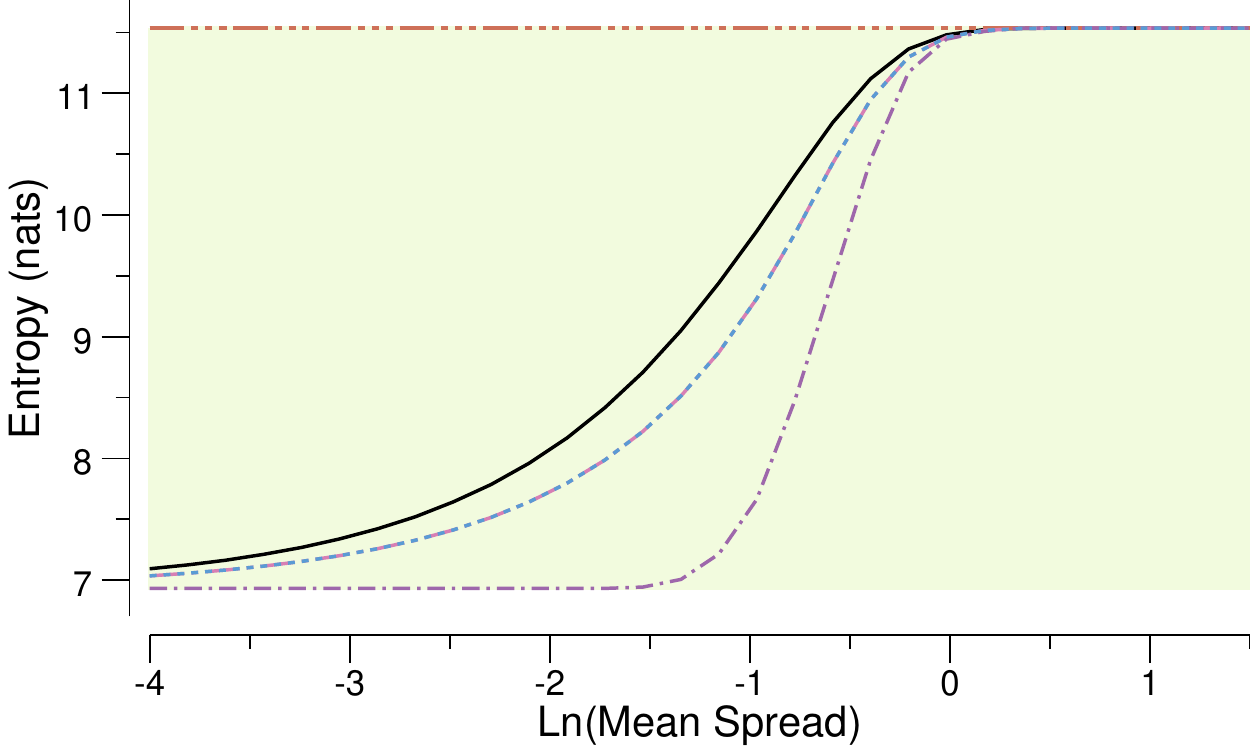}}{A}
\vspace{15pt}
\end{subfigure}
\begin{subfigure}[H]{0.49\linewidth}
\subplotlabel{

\includegraphics[width=\figwidth\linewidth]{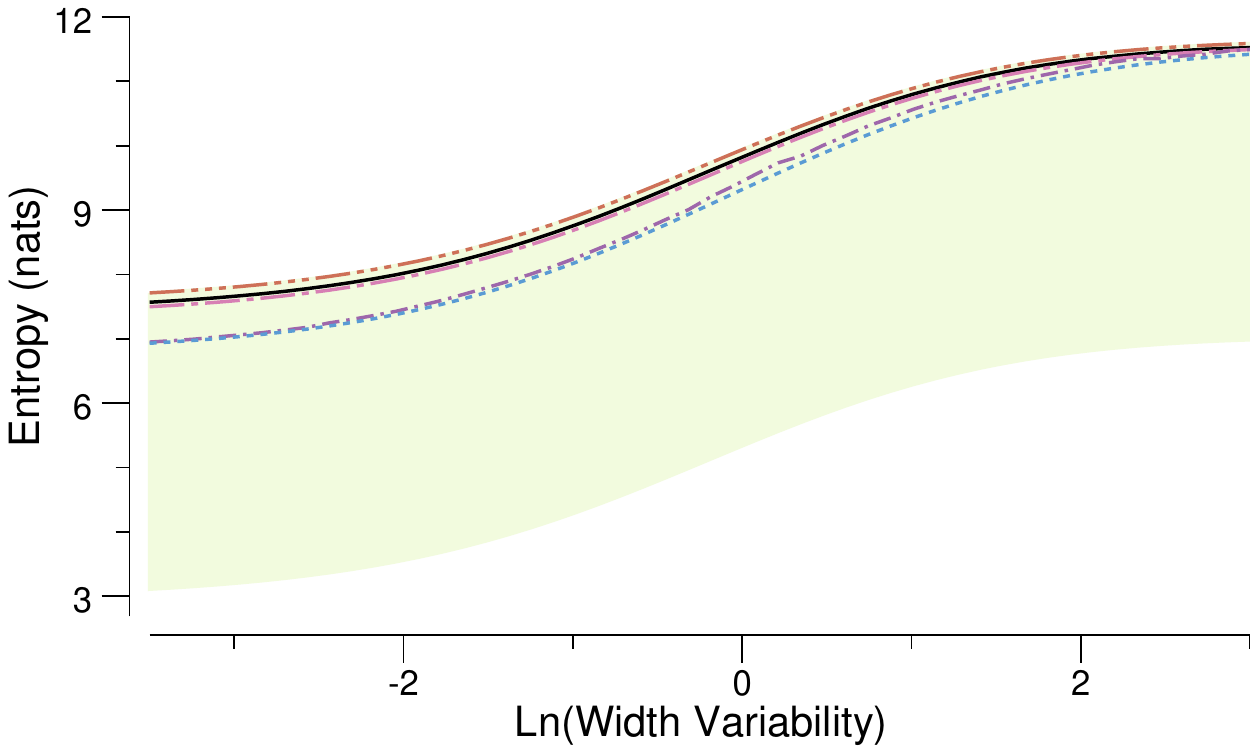}

}{B}\vspace{15pt}
\end{subfigure}\hfill
\\
\begin{subfigure}[H]{0.49\linewidth}
\subplotlabel{

\includegraphics[width=\figwidth\linewidth]{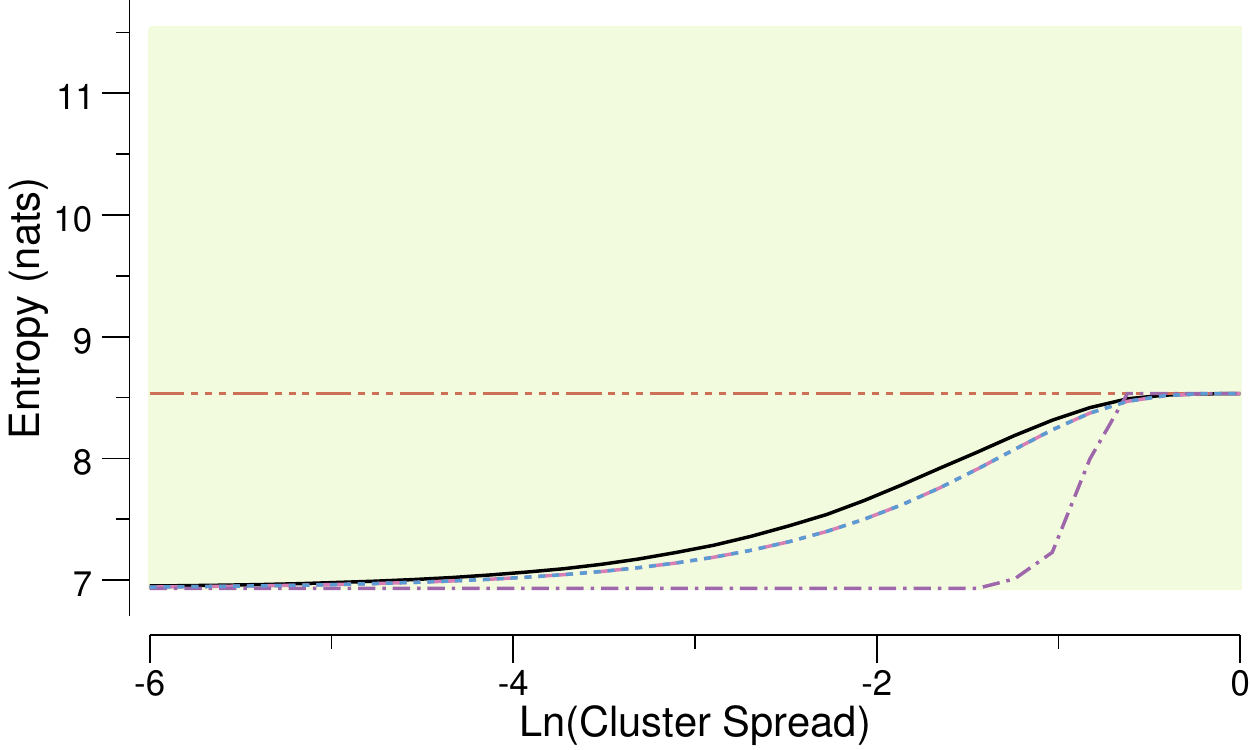}

}{C}
\end{subfigure}
\begin{subfigure}[H]{0.49\linewidth}
\subplotlabel{
\begin{raggedright}
\includegraphics[width=\figwidth\linewidth]{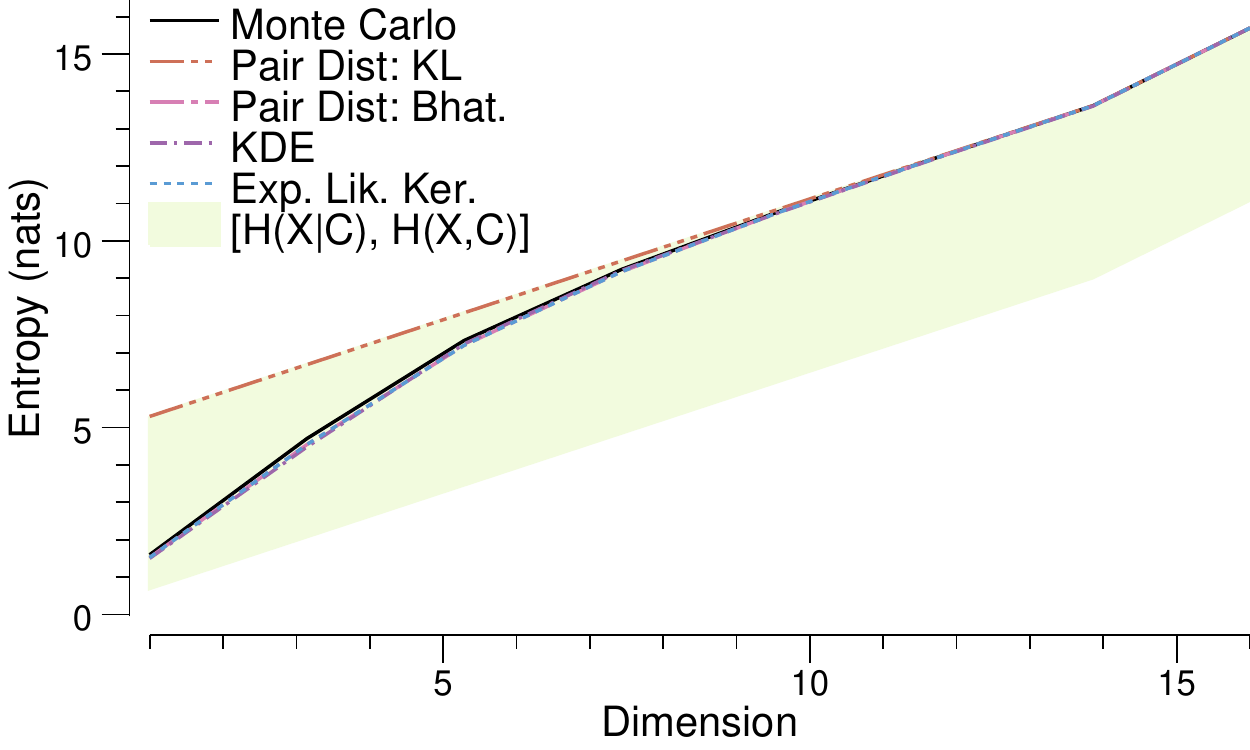}
\par\end{raggedright}
}{D}
\end{subfigure}

\caption{Entropy estimates for a mixture of a 100 uniform components. In each
plot, the vertical axis shows the entropy of the distribution, and
the horizontal axis changes a feature of the components: 
(\textbf{A})
the distance between means is increased;
(\textbf{B}) the component sizes become more similar (at the right side of the plot,
all components have approximately the same size);
(\textbf{C})
the components are grouped into five ``clusters'', and the distance
between these clusters is increased;
(\textbf{D})
the dimension is increased.\label{fig:centerspread-uniform}}
\end{figure*}

\subsection{Mixture of Uniforms}

In the second set of experiments, we consider a mixture of uniform
distributions. Unlike Gaussians, uniform distributions are bounded
within a hyper-rectangle and do not have full support over the domain.
In particular, a uniform distribution $p = \mathcal{U}({a}, {b})$ over $d$ dimensions
is defined as
$$
p(x) \propto \begin{cases}
	1, & \text{if } x_i \in [{a}_i, {b}_i] \; \forall i=1\dots d, \\
	0, & \text{otherwise,}
\end{cases}\,.
$$
where $x$, $a$, and $b$ are $d$-dimensional vectors, and the subscript $x_i$ refers to value of
$x$ on dimension $i$.
Note that when $p_X$ is a mixture of uniforms, there can be significant
regions where $p_{X}(x)>0$, but $p_{i}(x)=0$ for some $i$. 

Here, we list the formulae for pairwise distance measure between uniform
distributions. In the following, we use $V_{i} := \int_{x}1\{p_{i}(x)>0\} dx$ to indicate the ``volume''
of distribution $p_{i}$. Uniform components have a constant $p(x)$
over their support, and so $p_{i}(x) = 1/V_{i}$ for all $x$ where
$p_{i}(x)>0$. Similarly, we use $V_{i\cap j}$ as the ``volume
of overlap'' between $p_{i}$ and $p_{j}$, i.e., the volume of the
intersection of the support of $p_{i}$ and $p_{j}$, $V_{i\cap j} := \int_{x}1\{p_{i}(x)>0\}1\{p_{j}(x)>0\}dx$.
The distance measures between uniforms are then
\begin{align}
H(p_{i})  & =  \ln V_{i}, \nonumber \\
\text{KL}(p_{i}\Vert p_{j})  & = \begin{cases}
\ln(V_{j}/V_{i}), & \text{supp }p_{i}\subseteq\text{supp }p_{j},\\
\infty, & \text{otherwise,}
\end{cases} \nonumber \\
\text{BD}(p_{i}||p_{j})  &  =  0.5\ln V_{i}+0.5\ln V_{j}-\ln V_{i\cap j}, \label{eq:unif-bd} \\
\hat{H}_{\text{ELK}}  &  =  -\sum_{i}c_{i}\ln\sum_{j}c_{j}\frac{V_{i\cap j}}{V_{i}V_{j}} \label{eq:unif-elk} \,.
\end{align}

Like the Gaussian case, we run four different computational experiments
and compare the mixture entropy estimates to the true entropy, as
determined by Monte Carlo sampling.

In the first experiment, the mixture consists of $100$ 10-dimensional
uniform components, with $p_{i} = \mathcal{{U}}(\mu_{i}-\bm{1}_{(10)},\mu_{i}+\bm{1}_{(10)})$
, and $\mu_{i}\sim \mathcal{{N}}(0,\sigma\bm{I}_{(10)})$,
where $\bm{1}_{(d)}$ refers to a $d$-dimensional vector of $1$s.
Figure \ref{fig:centerspread-uniform}A
depicts the change in entropy as a function of $\ln(\sigma)$.
For very small $\sigma$, the distributions are almost entirely overlapping,
while for large $\sigma$ they tend very far apart. As expected, the
entropy increases with $\sigma$. Here, we see that the prediction
of $\hat{{H}}_{\text{KL}}$ is identical to $H(X,C)$, which arises
because $\text{KL}(p_{i}\Vert p_{j})$ is infinite whenever the support of $p_{i}$
is not entirely contained in the support of $p_{j}$. Uniform components with equal size
and non-equal means must have some region of non-overlap, and so the
$\text{KL}$ is infinite between all pairs of components, thus KL
is effectively $D_{\text{max}}$ (Equation (\ref{eq:dmin-dmax})). In contrast,
we see that $\hat{{H}}_{\text{BD}}$ estimates the true entropy quite
well. This example demonstrates that getting an accurate estimate
of mixture entropy may require selecting a distance function that
works will with the component distributions. Finally, it turns out
that, for uniform components of equal size, $\hat{{H}}_{\text{ELK}}=\hat{{H}}_{\text{BD}}$.
This can be seen by combining Equations (\ref{eq:pairwise}) and (\ref{eq:unif-bd}), and comparing 
to Equation (\ref{eq:unif-elk}) (note that $V_{i}=V_{j}$ when the components have equal size).

In the second experiment, we adjust the variance in the size of the uniform components.
We again use 100 10-dimensional components,
 $p_{i} = \mathcal{{U}}(\mu_{i}-\gamma_{i}\bm{1}_{(10)},\mu_{i}+\gamma_{i}\bm{1}_{(10)})$, where $\mu_{i}\sim \mathcal{{N}}(0,\bm{I}_{(10)})$, and $\gamma_{i}\sim\Gamma(1+\sigma,1+\sigma)$,
where $\Gamma(\alpha, \beta)$ is the Gamma distribution with shape parameter $\alpha$ and rate parameter $\beta$.  Figure \ref{fig:centerspread-uniform}B
shows the change in entropy estimates as a function of $\ln(\sigma)$. When $\sigma$
is small, the sizes have significant spread, while as $\sigma$ grows
the distributions become close to equally sized. We again see that
$\hat{{H}}_{\text{BD}}$ is a good estimator of entropy, outperforming
all of the other estimators. Generally, not all supports
will be non-overlapping, so $\hat{H}_{\text{KL}}$ will not necessarily
be equal to $H(X,C)$, though we find the two to be numerically quite
close. In this experiment, we find that the lower and upper bounds
specified by 
 $\hat{H}_{\text{BD}}$ and $\hat{{H}}_{\text{KL}}$ provide
a tight estimate of the true entropy.

In the third experiment, we again consider a clustered mixture, and evaluate the entropy estimators as these clusters
grow apart. Here, there are 100 components with 
{$p_{i} = \mathcal{{U}}(\tilde{\mu}_{g(i)}-\bm{1}_{(10)},\tilde{\mu}_{g(i)}+\bm{1}_{(10)})$,}
where $g(i)\in\left\{ 1 \dots 5\right\} $ is the randomly assigned cluster identity of component $i$. The cluster centers $\tilde{\mu}_{k}$ for $k\in\left\{ 1 \dots 5\right\}$
are generated according to $\mathcal{{N}}(0,\sigma\bm{I}_{(10)})$. Figure~\ref{fig:centerspread-uniform}C shows the change in entropy as the clusters locations move apart.
Note that, in this case, the upper bound $\hat{{H}}_{\text{KL}}$ significantly
outperforms $H(X,C)$, unlike in the first and second experiment, because in this experiment,
components in the same cluster have perfect overlap. We again see that $\hat{{H}}_{\text{BD}}$ provides a relatively accurate
lower bound for the true entropy.

In the final experiment, the dimension of the components is
varied. There are again 100 components, with $p_{i} = \mathcal{{U}}(\mu_{i}-\bm{1}_{(d)},\mu_{i}+\bm{1}_{(d)})$
, and $\mu_{i}\sim \mathcal{{N}}(0,\sigma\bm{I}_{(d)})$. Figure \ref{fig:centerspread-uniform}D
shows the change in entropy as the dimension increases from $d=1$ to $d=16$.  Interestingly, in the low-dimensional case, $H(X\vert C)$
is a very close estimate for the true entropy, while in the high-dimensional
case, the entropy becomes very close to $H(X,C)$. This is because
in higher dimensions, there is more `space' for the components to be far
from each other. As in the first experiment, $\hat{H}_{\text{KL}}$is
equal to $H(X,C)$. We again observe that $\hat{{H}}_{\text{BD}}$
provides a tight lower bound on the mixture entropy, regardless of dimension.

\section{Discussion\label{sec:Discussion}}

We have presented a new class of estimators for the entropy of a mixture
distribution. We have shown that any estimator in this class has a
bounded estimation bias, and that this class includes useful lower
and upper bounds on the entropy of a mixture. Finally, we show that
these bounds become exact when mixture components are grouped into
well-separated clusters. 

Our derivation of the bounds make use of some existing results~\cite{haussler_mutual_1997,hershey_approximating_2007}.
However, to our knowledge, these results have not been previously
used to estimate mixture entropies. Furthermore, they have not been
compared numerically or analytically to better-known bounds.

We evaluated these estimators using numerical simulations of mixtures
of Gaussians as well as mixtures of bounded (hypercube) uniform distributions.
Our results demonstrate that our estimators perform much better than
existing well-known estimators.

This estimator class can be especially useful for optimization problems
that involve minimization of entropy or mutual information. If the
distance function used in the pairwise estimator class is continuous
and smooth in the parameters of the mixture components, then the entropy
estimate is also continuous and smooth. This permits our estimators to be used within gradient-based
optimization techniques, for example gradient descent, as often done in machine learning problems.

In fact, we have used our upper bound to implement a non-parametric,
nonlinear version of the ``Information Bottleneck''~\cite{tishby_information_1999}.
Specifically, we minimized an upper bound on the mutual information
between input and hidden layer in a neural networks~\cite{kolchinsky_nonlinear_2017}.
We found that the optimal distributions were often clustered (Section \ref{subsec:clustered}).
That work demonstrated practically the value of having an accurate,
differentiable upper bound on mixture entropy that performs well in
the clustered regime.

Note that we have not proved that the bounds derived here are 
the best possible.  Identifying better bounds, or proving that our results are
optimal within some class of bounds, remains for future work.

\vspace{6pt}
\section*{Acknowledgments}
We thank David H. Wolpert for useful discussions. We would also like
to thank the Santa Fe Institute for helping to support this research.
This work was made possible through the support of AFOSR MURI on multi-information
sources of multi-physics systems under Award Number FA9550-15-1-0038.

\section*{Author Contributions}
Artemy Kolchinsky and Brendan D. Tracey designed the method and experiments; Brendan D. Tracey performed the experiments;  Artemy Kolchinsky and Brendan D. Tracey wrote the paper.

\section*{Conflicts of Interest}
The authors declare no conflict of interest.

\appendix

\section{Chernoff \boldmath{$\alpha$}-Divergence Is Not a Distance Function for \boldmath{$\alpha\protect\notin\left[0,1\right]$}
\label{appendix:chernoff-alpha-bounds}}

For any pair of densities $p$ and $q$, consider the Chernoff $\alpha$-divergence
\[
C_{\alpha}(p\Vert q):=-\ln\int p^{\alpha}(x)q^{1-\alpha}(x)\,dx=-\ln c_{\alpha}(p\Vert q),
\]
where the quantity $c_{\alpha}$ is called the Chernoff $\alpha$-coefficient
\cite{nielsen_chernoff_2011}. Taking the second derivative of $c_{\alpha}$ with
respect to $\alpha$ gives
\[
\frac{d^{2}}{d\alpha^{2}}c_{\alpha}(p\Vert q) =\int p^{\alpha}(x)q^{1-\alpha}(x)\left(\ln\frac{p(x)}{q(x)}\right)^{2}dx.
\]

Observe that this quantity is everywhere positive, meaning that $c_{\alpha}(p\Vert q)$
is convex everywhere. For simplicity, consider the case $p\ne q$,
in which case this function is \emph{strictly} convex. 
In addition, observe that for any $p$ and $q$, $c_\alpha(p\Vert q) = 1$ when $\alpha=0$ and $\alpha=1$.
If $c_\alpha(p\Vert q)$ is strictly convex in $\alpha$, this must mean that 
$c_{\alpha}(p\Vert q)>1$ for $\alpha\notin\left[0,1\right]$. This in turn
implies that the Chernoff $\alpha$-divergence $C_{\alpha}$ is strictly
negative for $\alpha\notin\left[0,1\right]$. Thus, $C_{\alpha}$
is not a valid distance function for $\alpha\notin\left[0,1\right]$,
as defined in Section \ref{sec:Pairwise-Estimators}.

\section{For Clustered Mixtures, \boldmath{$\hat{H}_{\text{BD}} \ge \hat{H}_{\text{ELK}}$ \label{appendix:bd-better-ekl}}}

Assume a mixture with perfect
clustering (Section \ref{subsec:clustered}). Specifically, we assume that if $g(i) = g(j)$, then  {$p_i(x) = p_j(x)$}, and if $g(i) \ne g(j)$ then both 
$e^{-C_{\alpha}(p_i || p_j)} \approx 0$ and $\int p_i(x) p_j(x) dx \approx 0$. 

In this case, our lower bound $\hat{H}_{\text{BD}}$ is at least as good as $\hat{H}_{\text{ELK}}$.
Specifically, $\hat{H}_\text{ELK}$ becomes
\begin{align*}
\hat{H}_\text{ELK} &= -\sum_i c_i \ln \sum_j c_j \int p_i(x) p_j(x) dx \\
&= -\sum_i c_i \ln \sum_j c_j \delta_{g(i), g(j)}  \int p_i(x)^2 dx \\
&= -\sum_k p_G(k) \ln \sum_k p_G(k)  \int p_k(x)^2 dx \,,
\end{align*}
where $p_k(x)$ is shorthand for the density of any component in cluster $k$ (remember that all components in the same cluster have equal density). 
$\hat{H}_{C_\alpha}$ becomes
\begin{align*}
\hat{H}_{C_\alpha} &= H(X|C) - \sum_i c_i \ln \sum_j c_j e^{-C_\alpha(p_i || p_j)} \\
&= -\sum_i c_i \int p_i(x) \ln p_i(x) dx - \sum_i c_i \ln \sum_j c_j \delta_{g(i), g(j)} \\
&= -\sum_k p_G(k) \int p_k(x) \ln p_k(x) dx - \sum_k p_G(k) \ln p_G(k) \\
&\stackrel{(a)}{\ge} -\sum_k p_G(k) \ln \int p_k(x)^2 dx - \sum_k p_G(k) \ln p_G(k) \\
&= -\sum_k p_G(k) \ln p_G(k) \int p_k(x)^2 dx = \hat{H}_\text{ELK} \,,
\end{align*}
where (\emph{a}) uses Jensen's inequality.

\ifdef{\articlenumber}{ 
}{\bibliographystyle{unsrt} }

\bibliography{bounds}

\end{document}